\documentclass{PoS}
\usepackage{enumitem}% http://ctan.org/pkg/enumitem
\usepackage{cite}
\usepackage{graphicx}
\usepackage[toc,page]{appendix}
\usepackage{url}
\usepackage{subcaption}
\title{An Automated Scalable Framework for Distributing Radio Astronomy Processing Across Clusters and Clouds}

\ShortTitle{Framework for distributing Radio Astronomy processing across Clusters and Clouds}

\author{\speaker{A.P. Mechev}\\
        Leiden University\\
        E-mail: \email{apmechev@strw.leidenuniv.nl}}

\author{J.B.R. Oonk\\
       Leiden University, ASTRON\\
       E-mail: \email{oonk@strw.leidenuniv.nl}}
       
\author{A. Danezi\\
       SURFsara\\
       E-mail: \email{anatoli.danezi@surfsara.nl}}

\author{T.W. Shimwell\\
       Leiden University\\
       E-mail: \email{shimwell@strw.leidenuniv.nl}}

\author{C.Schrijvers\\
       SURFsara\\
       E-mail: \email{coen.schrijvers@surfsara.nl}}
       
\author{H.T. Intema\\
       Leiden University\\
       E-mail: \email{intema@strw.leidenuniv.nl}}
       
\author{A. Plaat\\
       Leiden University\\
       E-mail: \email{a.plaat@liacs.leidenuniv.nl}}
       
\author{H.J.A. R\"{o}ttgering\\
       Leiden University\\
       E-mail: \email{rottgering@strw.leidenuniv.nl}}
       
\abstract{ The Low Frequency Array (LOFAR) radio telescope is an international aperture synthesis radio telescope used to study the Universe at low frequencies.  One of the goals of the LOFAR telescope is to conduct deep wide-field surveys. Here we will discuss a framework for the processing of the LOFAR Two Meter Sky Survey (LoTSS). This survey will produce close to 50 PB of data within five years. These data rates require processing at locations with high-speed access to the archived data.  

To complete the LoTSS project, the processing software needs to be made portable and moved to clusters with a high bandwidth connection to the data archive.  This work presents a framework that makes the LOFAR software portable, and is used to scale out LOFAR data reduction. Previous work was successful in pre-processing LOFAR data on a cluster of isolated nodes. This framework builds upon it and and is currently operational. It is designed to be portable, scalable, automated and general. This paper describes its design and high level operation and the initial results processing LoTSS data.}

\FullConference{International Symposium on Grids and Clouds 2017 -ISGC 2017-\\
		5-10 March 2017\\
		Academia Sinica, Taipei, Taiwan}

\begin{document}

\section{Introduction}
The LOFAR radio telescope is the world's largest aperture synthesis array with more than 20,000 antennas, and baselines of 60 m to 1000 km\cite{van2013lofar}. With its unprecedented sensitivity and angular resolution at ultra-low frequencies, LOFAR's goals are far reaching: from studying pulsars and supernova remnants in the Milky Way to the evolution of distant galaxies and the Epoch of Reionization\cite{EOR}. Additionally, LOFAR is a pathfinder for the larger Square Kilometer Array (SKA)\cite{SKA1} radio telescope. At low frequencies, the SKA telescope is expected to increase the data rate\cite{lofar_data} to 400TB per day creating more than 120PB per year\cite{ska_cloud_memo}. 

The LOFAR Two Meter Sky Survey (LoTSS) \cite{lotss} is observing 3000 different fields that will collectively map the entire northern radio sky. The majority of these datasets are anticipated to be 16TB per field. As such, the survey will create a total of 48 Petabytes. To complete the LoTSS survey in the project's target 5 year duration, $\sim$1PB of data must be processed each month. To mitigate delays caused by data transfer, processing must be done at a location with a high bandwidth connection to the raw data. SURFsara in Amsterdam is one such site and is also one of the LOFAR data archive locations.

Software packages for the initial processing of LOFAR data already exist\cite{van2016lofar}, however they were not implemented to efficiently operate on all cluster architectures. Environments with isolated compute nodes are a case where the current processing cannot fully use the resources. To complete the LoTSS project, a framework is needed to automatically process multiple datasets at the SURFsara location. This location has a large computing capacity, and has previous success with LOFAR processing\cite{oonk}.

We've built a framework on top of the LOFAR DSP\footnote{Distributed Shared Processing} platform \cite{oonk} named the
\textbf{LOFAR Reduction Tools} (LRT).The LRT software provides:

\begin{itemize}[noitemsep,topsep=0pt]
 \item Automation, enabling processing of multiple concurrent jobs
 \item Portability, enabling processing at different locations 
 \item Scalability, enabling adding worker machines as required by the workload
 \item Generalization, enabling integration of software from other scientific domains
\end{itemize}

In this paper, we present the implementation of the LOFAR processing pipeline for Direction Independent calibration, also known as `pre-FACTOR'\cite{prefactor},  into this framework. This software has been in use since November 2016 and at the time of writing (March 2017) has processed more than 100 datasets. This corresponds to a rate of roughly one dataset per day. By deploying the LRT framework on a cluster with a high-bandwidth connection to the data, the entirety of the LoTSS data can thus be reduced within the five year timespan of the project. 

The paper is structured as follows: Section \ref{sec:related} lists current work related to the research question. Section \ref{sec:lofar_red} outlines the LOFAR data reduction process and computational requirements. Section \ref{sec:design} describes the design of the LRT framework, its capabilities, the modification of the existing LOFAR software, performance and results. Finally, conclusions  and future work are discussed in Section \ref{sec:conclusion}.

\section{Related Work}\label{sec:related}

Scientific projects have begun producing petabyte-size datasets\cite{petabyte}. With increasing data sizes, researchers have begun focusing on scalable ways to parallelize their workflows. Because of this, processing is increasingly moving to grid- and cloud-based distributed computing facilities. From genetic sequencing \cite{scalable} and bioinformatics\cite{bioinfo} to neuroscience\cite{neurogrid} and ecology \cite{ecoinfo}, ever growing datasets have driven the development of distributed workflow systems in science\cite{pegasus}\cite{swift}. 

The framework presented in this publication is built on previous work distributing LOFAR pre-processing on a computing cluster\cite{oonk} using a PiCaS server\cite{picas} to track progress.  The details of this platform, the LOFAR DSP\cite{oonk}, are discussed elsewhere. Here we only provide a brief overview of the elements in this platform that interact with the LRT framework. The platform for LOFAR processing includes a PiCaS server and a CernVM Filesystem (CVMFS) client\cite{cvmfs2008}, previously deployed and tested on the target cluster. Additionally, continued technical support for this platform is provided by the SURFsara science support group. 

PiCaS is a token pool database used to create tokens describing processing jobs. It is built upon the CouchDB database\cite{couchdb}.  PiCaS\cite{picas_git} and CouchDB have been used in other distributed computing projects to launch and monitor jobs and store metadata. Job monitoring using PiCaS is also used in projects such as Sim-City\cite{simcity} and Finite Element modelling for sea dyke design\cite{li2017reliability}. In these works, pilot jobs were automatically launched and tracked remotely using the PiCaS software. 

CouchDB is also successfully used by the LHCb team to monitor the nightly software build process \cite{clemencic2014new} and by Sante et al.\cite{sante2010development} to launch asynchronous jobs to visualize and analyse gene sequencing data. As CouchDB documents can hold arbitrary information and attachments, CouchDB is suited for projects requiring the storing of metadata for many concurrent jobs, such as our application. 

CVMFS\cite{cvmfs2008} has been used by projects to package and publish software. The software used by many projects in High Energy physics, for example ATLAS \cite{cvmfsatlas} and the NO$\nu$A \cite{cvmfsnova} groups. These groups compiled scientific software on a central server and publish it to worker nodes. The LOFAR software has been similarly packaged\cite{oonk}. This makes deployment of processing scripts possible without a priori compilation on the distributed computing worker nodes. 

\section{LOFAR Data Processing}\label{sec:lofar_red}

Creating images from LOFAR data requires several steps of calibration and imaging. In order to place this work in the proper radio astronomy context, a brief introduction to the data processing in the context of the LoTSS is presented below. Section \ref{sec:image} gives an overview of LOFAR processing from an archived observation to a final image (Fig.\ref{fig:both_pipes}). Section \ref{sec:dirin_process} details the processing steps currently implemented as well as their computational challenges. Section \ref{sec:impl} contains an overview of the benefits of integrating the processing software with the LRT framework and a description of the processing by focusing on the dataflow (Fig. \ref{fig:DIpipe}).

\subsection{Producing Images From LOFAR Data}\label{sec:image}

The raw LoTSS data is stored at two locations of the LOFAR Long Term Archive\cite{LTA}. Typically each dataset is 16 TB and is split into 244 files of 65GB. Throughout the LoTSS data processing, this data is reduced to a $\sim$500GB set of calibrated data. The calibrated data is then imaged producing a final set of several 1.2GB images of 25k$\times$25k pixels each. The calibrated dataset is archived to allow for future refinement and re-imaging. 

\begin{figure}
\centering
 \includegraphics[width=.81\textwidth]{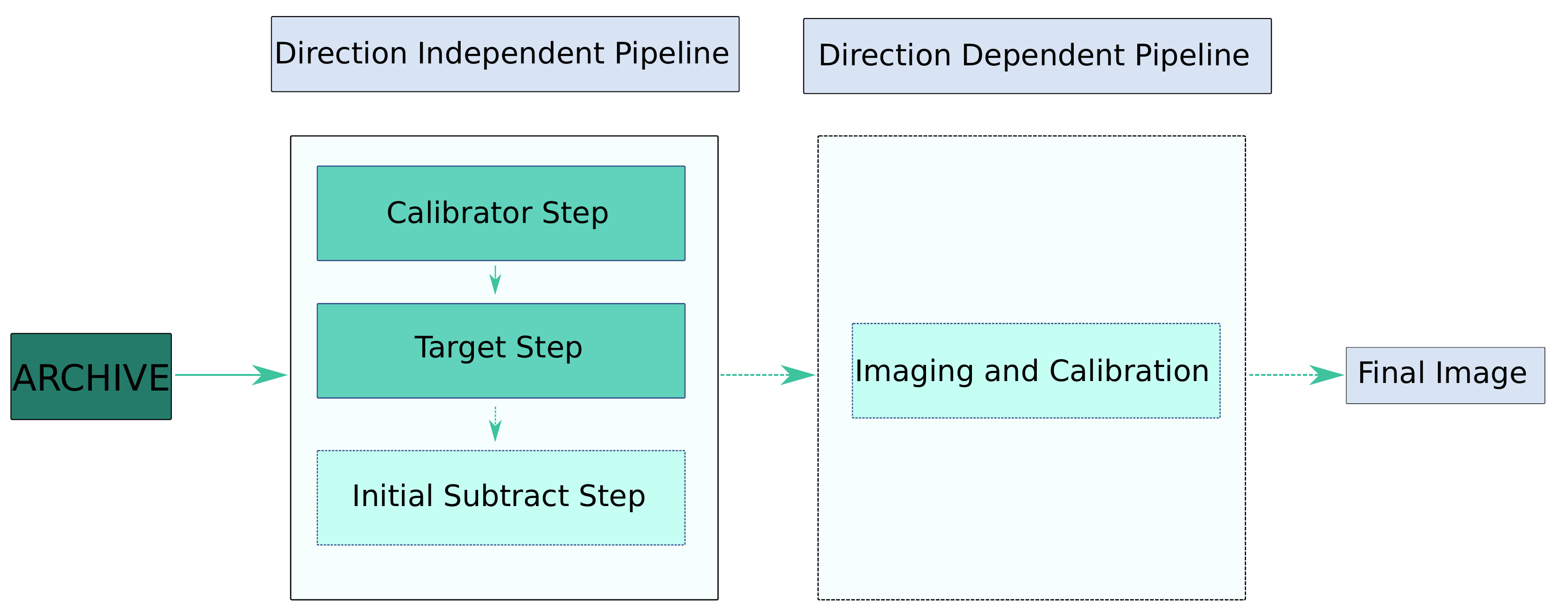}\\
 \caption{A schematic of the data flow through the Direction Independent Pipeline (pre-FACTOR\cite{prefactor}) and one of the Direction Dependent Pipelines, DDFacet (Tasse et al. in prep). The Initial Source Subtraction step is only necessary for some DD pipelines and is not currently implemented. }
 \label{fig:both_pipes}
\end{figure}

Data reduction for the LoTSS survey is split into two pipelines: Direction Independent (DI) pipeline followed by the Direction Dependent (DD) pipeline (Fig\ref{fig:both_pipes}). The Direction Independent pipeline\cite{van2016lofar}\cite{prefactor} is the first pipeline of LOFAR processing. It is necessary to produce a suitable starting point for the DD pipeline, however it produces images that are limited in resolution and contain residual instrumental effects\cite{lotss}\cite{van2016lofar} (Fig\ref{fig:artefacts}). In the DD pipeline, to achieve high fidelity continuum images, the ionospheric and beam errors must be corrected\cite{lofar_calib}. The latter effects vary across the field of view.  Here we present an implementation of the DI pipeline and the framework built to automate it. The LRT framework is built on top of the LOFAR DSP platform\cite{oonk} and runs on the Dutch GRID infrastructure. Implementation of the DD pipeline within the same framework is ongoing and will be presented in the future.

\subsection{Direction Independent Pipeline}\label{sec:dirin_process}

The LOFAR telescope consists of many stations (clusters of electronically coupled antennas), each with an independent electronic gain. The station-based gain calibration parameters can be deduced from the observation of a bright calibrator source before or after the science target\cite{lofar_calib}. This calculation is performed by the calibration step of the DI pipeline (Fig\ref{fig:both_pipes}).  The results from this step are applied to the science target, which is then averaged and processed. This includes removing Radio Frequency Interference and subtraction of bright off-axis sources, and finally calibration against a skymodel derived from other radio surveys\cite{lotss}\cite{van2016lofar}\cite{lofar_calib}. These steps are performed by the Target step of pre-FACTOR\cite{prefactor}. The result is a DI-calibrated dataset which is up to 64 times smaller than the uncalibrated archived data.

%The initial part of the LOFAR data processing can benefit from a cluster of many small machines. Doing so not only improves latency but allows for both scalability and fault tolerance, as discussed in Sections  \ref{sec:intermediate_storage} and \ref{sec:capabilities}. 

The Direction Independent pipeline (Fig.\ref{fig:DIpipe}) consists of an existing set of scripts which use the LOFAR software suite\cite{lofarcookbook} and pre-FACTOR\cite{prefactor} to process the archived data. A parameter-set file (parset) defines a sequence of procedures and their corresponding input parameters. Each procedure may launch one or more executables.

\subsection{DI Data Flow and GRID Implementation}\label{sec:impl}

The LoTSS survey is led by Leiden University and conducted by a large international group of scientists. Most of the host institutions of those scientists do not have a dedicated network connection to the LOFAR data archive. This means they must download archived (16TB per dataset) data  over a public connection. In the case of Leiden, the sustained download rate is 10 to 30 MB/s. This is too low to download a full-size archived dataset in a reasonable time. Downloading of one dataset completes in two weeks, 10 times longer than the DI processing. At this rate, transferring 3000 datasets would take up to a century. This download bottleneck was already recognized by the LOFAR spectroscopy group (PI Oonk) who developed the LOFAR DSP platform\cite{oonk} for large-scale processing. 

To mitigate download issues, the processing  was moved to the SURFsara compute grid location at the Amsterdam Science Park\footnote{\href{http://docs.surfsaralabs.nl/projects/grid/en/latest/Pages/Service/system_specs.html}{http://docs.surfsaralabs.nl/projects/grid/en/latest/Pages/Service/system\_specs.html}} as there were previous successes in processing LOFAR data at SURFsara by the LOFAR Spectroscopy group\cite{oonk}.  The pre-FACTOR package runs within the generic pipeline framework\footnote{\href{http://www.astron.nl/citt/genericpipeline/}{http://www.astron.nl/citt/genericpipeline/}} which is part of the LOFAR software stack\cite{lofarcookbook}. This framework cannot run on the SURFsara Gina cluster\cite{gina_specs} out of the box. It requires either a mounted shared file system or node to node communication, and the Gina nodes offer neither. Work was done to implement the current pre-FACTOR package on the existing LOFAR DSP platform \cite{oonk}. This work resulted in the LRT framework presented here: a package allowing the implementation of different LOFAR processing pipelines on a distributed infrastructure.

\begin{figure}
\centering
 \includegraphics[width=.89\textwidth]{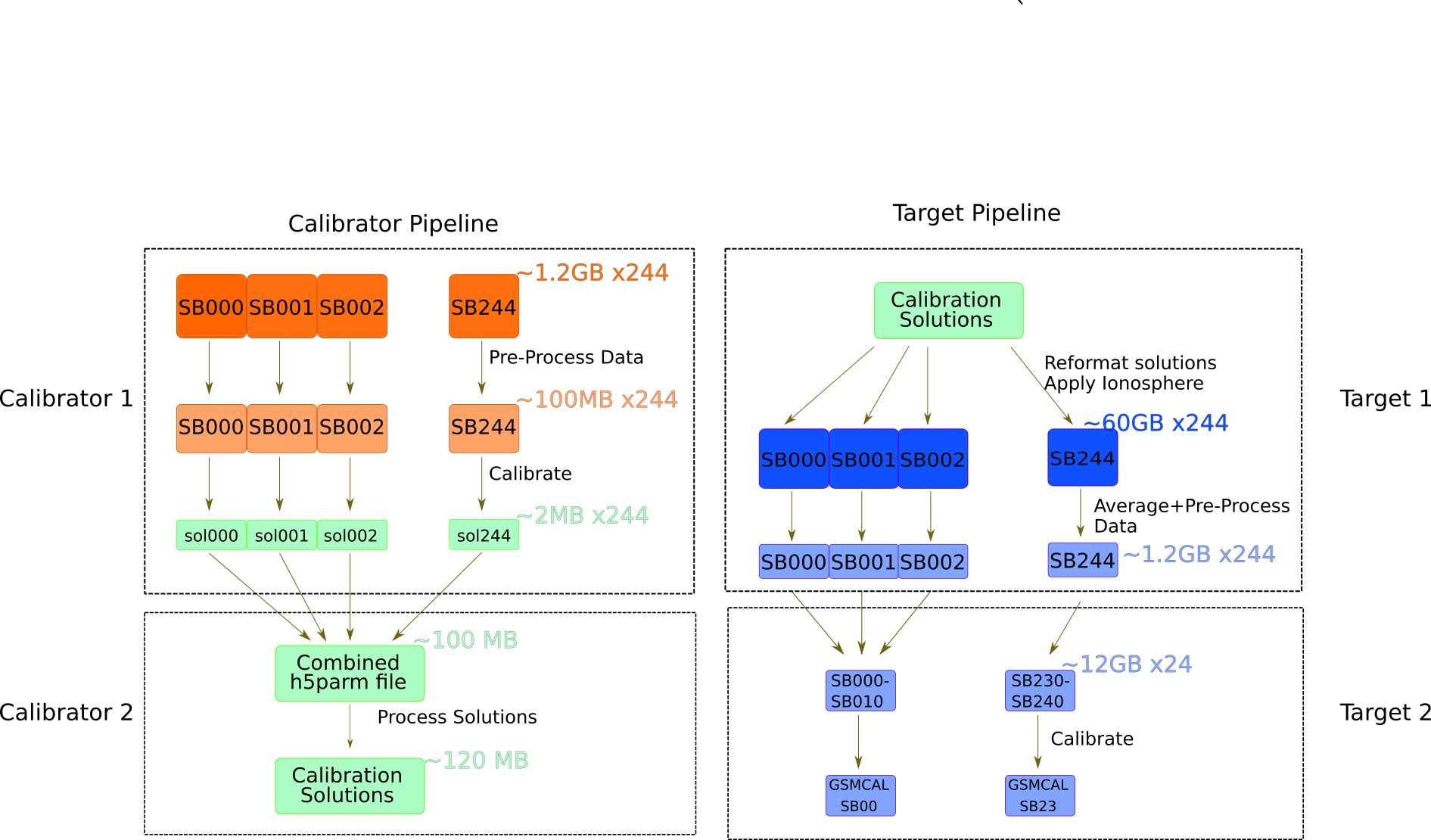}\\
 \caption{Data flow and parallelization of the Direction Independent Processing. The Calibrator 1 and Target 1 steps run concurrently as independent jobs. Calibrator 2 and Target 2 combine these results. Note that the Target 1 step requires the solutions produced by the Calibrator 2 step. This places a strict ordering on the processing steps. }
 \label{fig:DIpipe}
\end{figure}

In the case of pre-FACTOR, the two steps of the DI reduction, the Calibrator and Target, were each split in two parts (Fig.\ref{fig:DIpipe}). The first parts of the Calibrator and Target processing are parallelized by running one subband per node, resulting in 244 concurrent jobs. This takes advantage of the data level parallelism of initial LOFAR processing. Running 244 concurrent jobs is also a natural way to process the data as each observation is stored in 244 individual files (as discussed in Section \ref{sec:image}). 

Additionally, splitting the computation makes it more robust. In the case that the download or processing of one job fails, it can be restarted without disrupting parallel jobs. When a step has finished processing (for instance, the Calibration step in Fig.\ref{fig:both_pipes}), the next step can be launched automatically enabling the massive processing of LOFAR Surveys data. 

The pre-FACTOR software was designed to be run on single node or clusters with a shared file system. Because the worker nodes at the SURFsara cluster have isolated storage, scripts are included in the LOFAR Reduction Tools to load the relevant data on the worker node before processing. After a job is finished, the scripts save intermediate results to an external storage location\cite{oonk}.

Using intermediate storage to hold the results from each step, the pre-FACTOR DI pipeline was split into four steps as shown in Fig. \ref{fig:DIpipe}. The Calibrator 1 and Target 1 steps download the raw data at one file per worker node and store the processing results (Calibration Tables and Processed data respectively) in storage. After all Calibrator 1 jobs finish, the Calibrator 2 step combines the results produced into a single calibration table. This table is then applied to the science target by the Target 1 jobs. Finally the Target 2 job combines 10 datasets produced by Target 1 and creates the final DI calibrated datasets. 

% \subsection{Data Flow and Processing Steps}\label{sec:dataflow}
% 
% A researcher interested in processing LOFAR data needs to download it from the LOFAR Long Term Archive. Leiden University has a computer cluster dedicated for LOFAR processing. The connection between the Leiden location and the LOFAR data archive is typically 10 MB/s. 
% 
% Unlike at Leiden University, the SURFsara clusters have a gigabit connection to the data archive, accelerating the data retrieval to 1.5 days. Additionally, having two orders of magnitude more processing nodes than at Leiden, the reduction can be further parallelized. This has accelerated the processing of one (downloaded) dataset from more than two days to less than half a day.

\section{Framework Design}\label{sec:design}

\begin{figure}
\centering
 \includegraphics[width=0.8\textwidth]{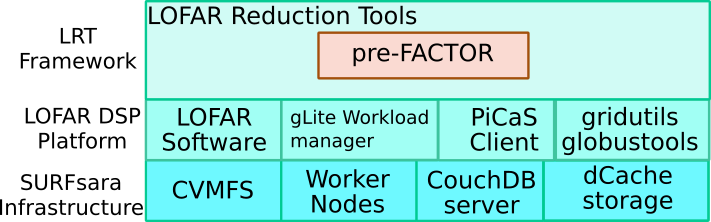}\\
 \caption{Schematic of the dependencies of the LRT framework, the modules of the LOFAR DSP platform\cite{oonk}, and the infrastructure provided by SURFsara. }
 \label{fig:platform}
\end{figure}

\begin{figure}
 \includegraphics[width=\textwidth]{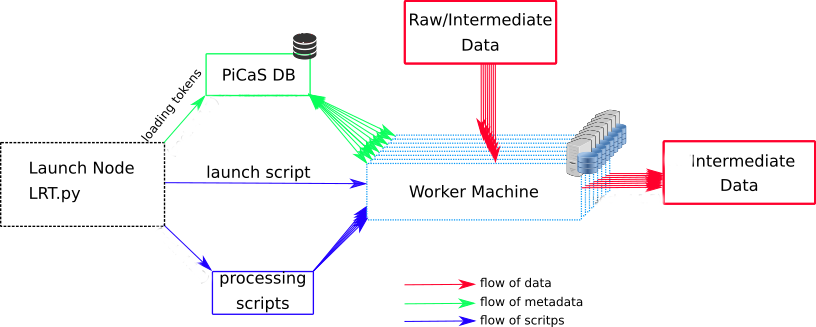}\\
 \caption{Overview of the design of the LRT framework. Shown is the decoupling of scripts from metadata, compute elements and data storage.}
 \label{fig:design}
\end{figure}

The LRT framework (Fig. \ref{fig:design}) was developed to automate the LOFAR Direction Independent calibration by processing the data on the Gina cluster at SURFsara\cite{gina_specs}. It is built on the LOFAR DSP platform\cite{oonk}, which facilitates building distributed computing frameworks. A component diagram is shown in Fig.\ref{fig:platform}.

% The goal of the framework was to split the pre-FACTOR processing to take advantage of the large computational resources and high bandwidth at this site. 

By making each job self-contained, the LRT framework provides a portable way to execute the LOFAR scripts (Section \ref{sec:port}). The Gina cluster provides more than 6000 cores over 300 processing nodes. To take advantage of these capabilities, the framework was made scalable (Section \ref{sec:scalable}). Typical processing regularly scales out to 244 jobs per step. To process the  LoTSS observations efficiently, automation was built into the LRT tools (Section \ref{sec:automation}). Finally, the framework is constructed to allow the inclusion of different processing suites, building onto the generality provided by the LOFAR DSP\cite{oonk} platform. This generality is now used by several1 LOFAR projects (Section \ref{sec:general}). 

\subsection{Framework Elements}
The LRT framework consists of a set of modules responsible for different parts of the data reduction. The \verb|srmlist| module handles the storage links to the archived data. If the data is on tape, it sends a command to transfer it to disk. The \verb|sandbox| module packs the processing scripts in an archive and uploads it to disk storage on a dCache system\footnote{\href{http://docs.surfsaralabs.nl/projects/grid/en/latest/Pages/Service/system_specifications/dcache_specs.html}{http://docs.surfsaralabs.nl/projects/grid/en/latest/Pages/Service/system\_specifications/dcache\_specs.html}
}\cite{dcache}. Doing so makes the processing portable. The \verb|Token| module is responsible for managing metadata, creating job tokens which define a processing job. As the tokens can store arbitrary data, the LRT modules are easily generalizable to other workflows. Documentation of these modules and examples can be found on the LRT github page\footnote{\href{https://github.com/apmechev/GRID_LRT}{https://github.com/apmechev/GRID\_LRT}}.

\subsection{Framework Capabilities}\label{sec:capabilities}

The LRT implementation is designed to be platform independent. It allows for easy extension, thereby enabling LOFAR reduction schemes other than the LoTSS reduction. Two examples of this are the updated LOFAR GRID spectroscopy and LOFAR GRID pre-processing pipelines\cite{oonk}. We will describe how we use the LOFAR DSP platform to achieve scalability, portability, automation and generality. 

\subsection{Portability}\label{sec:port}

\begin{figure}
 \includegraphics[width=.76\textwidth]{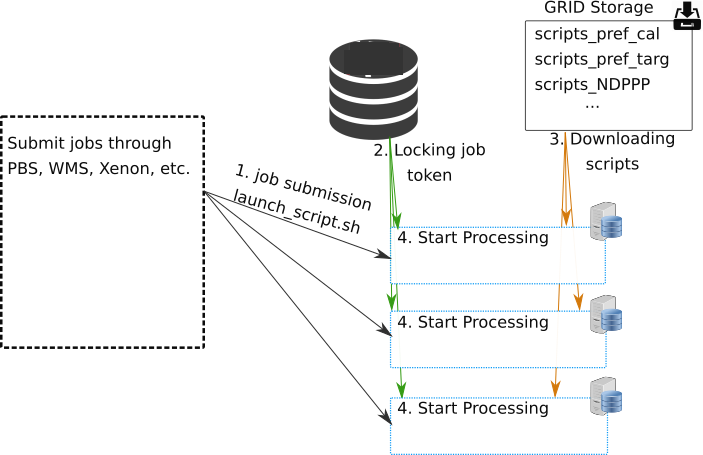}\\
 \caption{Starting processing on worker nodes. Currently processing is done on SURFsara Gina cluster , however the framework has been tested at the Leiden University cluster.}
 \label{fig:tok_run}
\end{figure}

The Gina architecture requires processing scripts to be stored remotely from worker nodes. These scripts are archived and uploaded to distributed storage and their location is added to the job description. After a worker node locks a job token, it downloads and extracts the processing scripts, reads the metadata from the token and begins the processing (Fig.\ref{fig:tok_run}).

Storing the scripts location in the job description allows different steps of the pipeline to use different sets of scripts. The benefit from this design is that as long as the worker node has access to the Universal Resource Identifier (URI) of the scripts, it can process the data. This choice provides portability, enabling processing over a variety of distributed computing environments, including sites in the the European Grid Infrastructure\cite{egi}.

The PiCaS tokens can store strings and integer values as well as file attachments. The LRT DI pipeline implementation uses attachments to store diagnostic files used to assess data quality, lists of links to the data, and parset files that define the pre-FACTOR workflow. Storing these files in a central database means any worker node can read this data at runtime, regardless of the node's location. 

The pre-FACTOR scripts require an installation of the LOFAR software stack\cite{lofar_stack}. These requirements are met by mounting a CVMFS\cite{cvmfs2008}\cite{softdrive} installation of the LOFAR software stack. The CVMFS service provides a portable pre-compiled copy of the LOFAR software. With the CVMFS prerequisite satisfied and an active grid proxy, any computer can download the data and participate in any data reduction step.

\subsection{Scalability}\label{sec:scalable}

\begin{figure}
 \includegraphics[width=.76\textwidth]{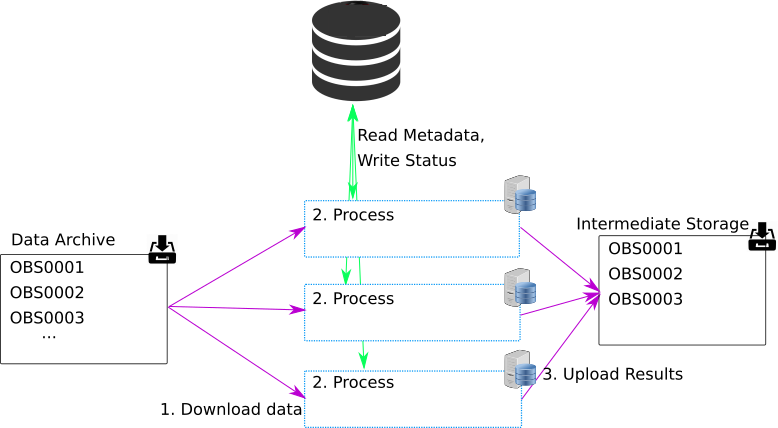}\\
 \caption{Processing of LOFAR data from the Long Term Archive with results stored at an intermediate storage location.}
 \label{fig:tok_process}
\end{figure}

The LRT framework can define, launch and monitor jobs on a distributed computing infrastructure. As such, it is effective for pipelines that independently process large datasets in parallel. Each part of the dataset is processed on a single node, and the metadata of this job is stored and updated in a remote database which can be read from and written to by the worker node. A schematic of the communication between worker nodes and the PiCaS database is in Fig.\ref{fig:tok_process}. 

By using a concurrent document-oriented database such as CouchDB\cite{couchdb}, each document can store the metadata describing a single processing job. This is not possible with relational databases such as MySQL\cite{yarger1999mysql}. These documents are called Tokens, as defined by the PiCaS framework\cite{picas}. Processing is scaled out by creating the required number of jobs and launching them on independent nodes. This system can easily scale to tens of thousands of jobs, and currently stores the metadata of thousands of LOFAR jobs. 

The first implementation of PiCaS and CouchDB in the LOFAR-DSP platform was carried out in the context of the LOFAR spectroscopy project and custom user processing\cite{oonk}. This first implementation focused on processing individual data sets and required a high-level of user interaction. Here we build upon the LOFAR DSP platform by making it easier to define the structure of tokens in a text file. Additionally, we provide the automation to process multiple runs (calibrator and target) and handle their products.

\subsection{Intermediate Data Storage}\label{sec:intermediate_storage}

Splitting the processing into multiple steps requires intermediate data to be stored at a location accessible to the worker nodes. As the current processing is done at the Gina cluster, the intermediate results are stored in several dedicated
storage pools hosted by SURFsara. At each step, the LRT processing scripts check whether the required input data exists and downloads it. This avoids unnecessary repetition of reduction steps and allows to restart processing from the point of failure rather than from the start. Since the PiCaS tokens can hold the location of the intermediate data, processing becomes scalable to any location that has access to the intermediate data. For example, the final DI datasets are used by DD calibration at several institutes in the Netherlands and Europe. 

\subsection{Automation}\label{sec:automation}

Running many jobs in parallel as part of a multi-step workflow requires automatically launching and restarting steps.
Because of the strict ordering of the pipeline steps, each step must wait for the previous step to complete. A PiCaS query is used to tally the number of completed jobs in a step. Once a threshold is reached, the next step automatically launches. Since PiCaS stores the state of each job, failed runs can be restarted automatically by a script which monitors the status of jobs in the database.  

Creating tokens is also automated. The user can define the structure of their job token in a text file, and use that file to automatically create tokens. This allows easy and fast creation of tokens holding different sets of metadata. Similarly, the scripts destined for a worker node are packed in an archive called a 'sandbox'. The list of scripts and repositories stored in this archive are stored in a text file, allowing to automatically create different sandboxes by changing this configuration file. A user only has to specify the ID of the observation they're interested in, and the list of files they need to process before launching the DI pipeline.

\subsection{Generality}\label{sec:general}

A PiCaS token can hold arbitrary metadata and store configuration files. A user can define their workflow by deciding on the set of steps and the processing done at each step. Once each step launches, it can read the metadata it requires from the Token and load the required software from the CVMFS server described in Section \ref{sec:port}. Other pipelines can be combined with the LRT framework. This is done by defining the necessary token fields and specifying the processing scripts of the pipeline steps. 

While this work discusses the implementation of the LOFAR DI pipeline, work is ongoing to also port the LOFAR Direction Dependent pipeline on the LRT framework.

% Since the data location is static, the paths of the data are hard coded into the scripts. The framework design, however, also allows a job to log the location of its output data into the CouchDB database. Jobs in subsequent steps can read the location of their input data from the tokens of the previous job. This will be implemented when the LOFAR data processing becomes distributed across multiple locations. 

\subsection{LOFAR LoTSS Use Case}\label{sec:use}
% The LoTSS Survey requires processing of more than 8 PB of data each year in order to keep up with the data produced by the telescope. As there are more than 3000 observations planned, processing them manually is untenable. Additionally, the large raw data sizes require the data be reduced in parallel before the Direction Dependent calibration step since the data will not fit in the memory of a single machine. 

LOFAR observations are split and stored in frequency chunks called subbands. A LoTSS observation consists of 244 subbands  for the calibrator and 244 subbands for the target. The Calibrator 1 and Target 1 step of the DI Pipeline processes these subbands independently. This is a form of data-level parallelism and increases data throughput. 

Without a framework to automate and distribute the processing and a cluster at an LTA location, these datasets would need to be downloaded to an institute's cluster. Such standalone runs of the pre-FACTOR scripts typically process one observation in two weeks dominated by the data transfer time.  At the 10-30MB/s connection (the sustained speed at Leiden University), the downloading would take between 30 and 100 years. At SURFsara, the 1Gbps external connection is fast enough to download and process the 16TB in a day and a half. While clusters at typical institutions number in the tens of nodes, the  Gina cluster at SURFsara has over 300 nodes. Each of the 244 subbands is run on a dedicated node concurrently. This massive parallelization further increases data throughput. 

Porting the LOFAR LoTSS data reduction to the SURFsara Gina cluster using the LRT package has resulted in a 15x increase in throughput. Suggestions on further increasing the amount of data processed are presented in Section \ref{sec:future}.

% \subsection{Initial Results}\label{sec:performance_results}

%%%%%%%%%%MAKE ABOVE MORE CONCISE

% \subsection{Bottlenecks and Proposed Solutions}
% 
% Launching multiple processing jobs leads to large data transfer rates between the raw data archive and the processing site. While the SURFsara Grid location has a gigabit connection to the data archives, launching more than two reductions at a time results in a download bottleneck. This bottleneck is caused by the 1 Gbps public connection between LTA sites. Most of the LoTSS data is stored at the Forschungszentrum Jülich location. Thus, the 1Gbps bandwidth between the archive and processing sites limits the download to 35 hours per 16TB observation. To process the outstanding data within the five project time line, the minimum bandwidth requirement is a 2.5 Gbps uninterrupted connection to the data. Solutions to this bottleneck are presented in Section \ref{sec:future}. 

\section{Conclusion and Future Work}\label{sec:conclusion}

The goal of the LRT framework is to build upon the LOFAR DSP platform to create a package to port LOFAR processing to a massively distributed compute environment. The LRT framework was designed to be scalable, portable, automated and general. The DI pipeline of the LoTSS survey was used as a demonstration of the capabilities of the LRT software. 

Combining the DI pipeline scripts with the LRT tools resulted in a 15x increase in throughput compared to previous LoTSS data reduction strategies which were dominated by data transfer. Automation was provided by separating the different parts of execution into separate modules and using configuration files to facilitate creating workflows. Thanks to this automation, it is possible to perform the processing necessary for the LoTSS survey.

The portability of the LRT framework makes it easy to move processing to other compute locations such as those near the data archive, increasing the throughput. This portability is provided by installing the software on a CVMFS server that worker nodes can access and by storing metadata on an external PiCaS server. Separating scripts and metadata from the computation elements makes it possible to use computational resources at multiple sites as required.

The scalability of this framework allows to launch multiple data reductions concurrently and easily monitor progress through a web-accessible CouchDB interface. Scalability is achieved by storing metadata in a document based database with asynchronous write support, and by running each job on an isolated node. 

The framework is made general by using PiCaS tokens, capable of storing arbitrary metadata, and passing this data to the processing scripts. Additionally, by separating the processing from the data retrieval, the framework can `plug-and-play' different software and execute it on the same dataset. Finally, as the framework is general, other LOFAR projects can benefit from incorporating their processing into the LRT framework.

Using the LRT framework, more than 100 datasets have passed through the Direction Independent pipeline. This corresponds to a throughput of $\sim$1 dataset per day. Future improvements (Section \ref{sec:future}) are expected to increase the throughput to two datasets per day. This is the minimum rate required by the LoTSS survey.

% Currently, the LRT data reduction is launched manually. There are upcoming plans to create a trigger launched by the Observatory at the end of a successful observation. Using this trigger, the processing can be integrated with the data acquisition and launched automatically at the end of the observation. Doing so enables producing an image less than a week after the observation has completed without requiring human interaction. 
% 
\subsection{Future Work}\label{sec:future}

\begin{figure}[b!]

\begin{subfigure}[b]{\textwidth}
\centering
\includegraphics[width=.95\textwidth]{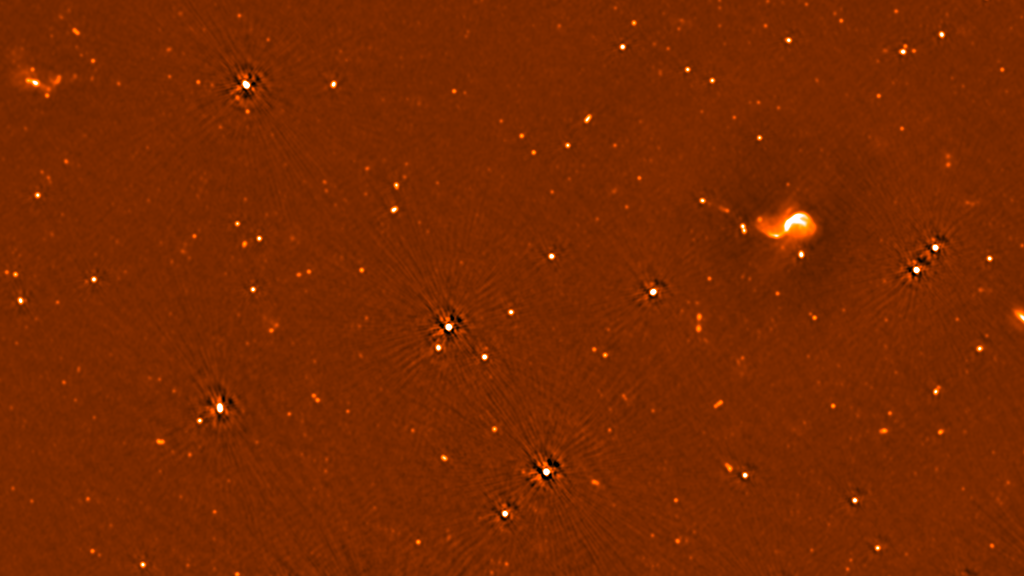}
\caption{Field centred on 12:15:00,+47:00:00 showing M106 on the right. }
\end{subfigure}%
\\
\begin{subfigure}[b]{\textwidth}
\centering
\includegraphics[width=.95\textwidth]{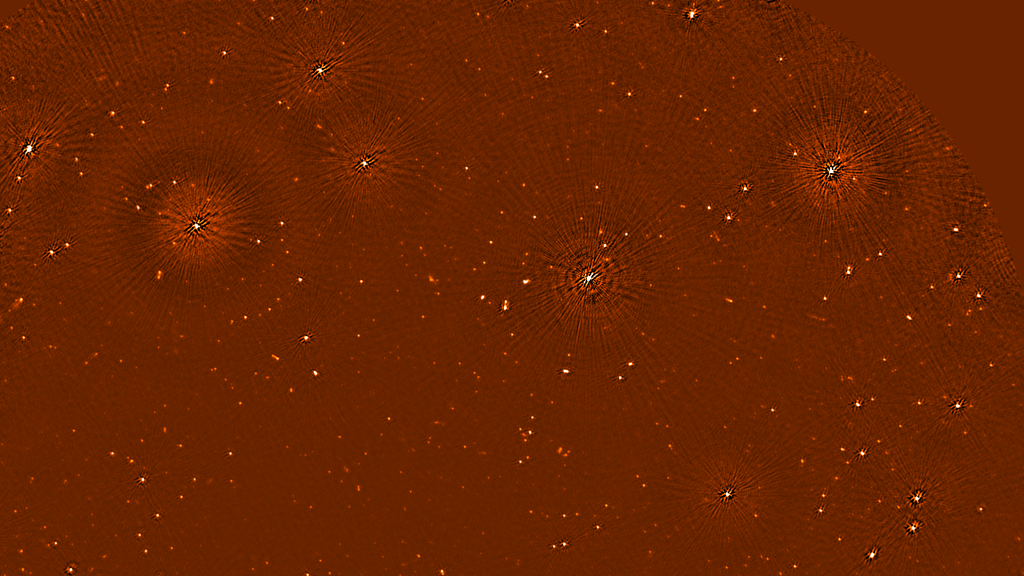}
\caption{Image of a field centred on 12:22:00,+53:40:00 produced by the DI pipeline. Artefacts around the bright point sources are to be removed by the DD pipeline. }
\label{fig:artefacts}
\end{subfigure}%

\caption{A sample of LoTSS image, after Direction Independent calibration with pre-FACTOR using the LRT framework. Artifacts such those in Fig. \ref{fig:artefacts} need to be removed by Direction Dependent processing (Section \ref{sec:image} , Figure \ref{fig:both_pipes})  }
\label{fig:galaxies}
\end{figure}

A significant portion of the LoTSS data is stored at the FZ J\"{u}lich data centre\footnote{\href{http://www.fz-juelich.de/portal/EN/Home/home_node.html}{http://www.fz-juelich.de/portal/EN/Home/home\_node.html}}. Because of the high data sizes, even the 1 Gbps transfer between this site and SURFsara is insufficient to process two observations per day. We are currently investigating porting the LRT framework to the FZJ site as well. Doing so will reduce the data size by a factor of 64. This will make transfer to other processing locations possible within a few hours for each dataset.

While the LRT framework successfully automated the Direction Independent calibration pipeline, it still needs to  implement the Direction Dependent processing scripts.  

% The Direction Dependent processing cannot be parallelized across nodes yet. This part of the processing currently takes more than four days per dataset per node. It is anticipated that the DD pipeline will be split into two steps. 

DIRAC\cite{dirac} or Xenon\cite{maasen_xenon} are two middleware packages that allow launching jobs at multiple clusters from a single location. DIRAC is expected to replace the current workload management system at EGI sites, and Xenon is used by some projects at SURFsara\cite{simcity}. Due to the portability of the LRT software, the LOFAR processing can be launched on other clusters using such middleware. Launching LRT jobs at other LTA sites will reduce the size of archived data so it can be transported faster. 

Finally, as the reduction is automated, it can in principle be started shortly after the telescope finishes the observation. Launching jobs immediately after an observation will minimize the overhead spent moving the data from tape to disk, which can take up to a week. 

Minimizing the latency between observation and science quality images will benefit the LOFAR community immensely. This fast turnover will allow radio astronomers to focus on their specific science case. An all-sky survey at the 150 MHz range will create a multitude of targets for follow-up with optical telescopes such as the LOFAR-WEAVE survey\cite{weave}. Figure \ref{fig:galaxies} shows a small sample of interesting objects in the data processed using the software presented in this publication.  A full list of science results expected from the LoTSS project can be found in\cite{lotss}. Efficient high-throughput processing of LOFAR data will empower these science cases opening the way to exciting new discoveries.

\section*{Acknowledgement}
This work was carried out on the Dutch national e-infrastructure with the support of SURF Cooperative through grant e-infra 160022 \& 160152.

This work was done with the support from the NWO/DOME/IBM programme ``Big Bang Big Data: Innovating ICT as a Driver For Astronomy'', project \#628.002.001

% \begin{appendices}
% 
% \section{Execution of LOFAR Reduction Tools }\label{sec:appendix_1}
% 
% The LRT framework handles staging of LOFAR data, packaging and uploading worker node scripts (named sandboxes), creating PiCaS job tokens and launching pilot jobs on the SURFsara Gina cluster. The framework is modular, allowing an user to execute any of the previous steps manually, or alternatively launch an automated reduction. It can be downloaded from the Github page (\href{https://github.com/apmechev/GRID_LRT}{github.com/apmechev/GRID\_LRT}) and installed with \\
% \verb|python setup.py build && python setup.py install| . Documentation on the usage of the tools can be found at the Github page. 
% \end{appendices}

\end{document}